# A new type nuclear reaction on $^{159}$Tb in the outgoing channel considering observation of a bound dineutron

I.M. Kadenko*

*International Nuclear Safety Center of Ukraine; Department of Nuclear Physics  
Taras Shevchenko National University of Kyiv, 01601, Kyiv, Ukraine*

A new type nuclear reaction on $^{159}$Tb with neutrons and protons in the incident channels and a bound dineutron ($^{2}n$) in the output channel is considered based on available experimental observations. The dineutron is assumed to be separated from the volume but not from the potential well of the residual nucleus. Such configuration represents a nuclear system with a satellite dineutron located at few fm distances from the surface of the residual nucleus. Due to dineutron disintegration the decay products may react with the residual nuclei, leading to their transformations much faster than expected

*Introduction.* - The purpose of this letter is discussion of the dineutron as a bound particle in the output channel near the heavy nucleus of certain nuclear reactions. This arrangement may represent a nuclear system consisting of the residual nucleus and a satellite particle (the dineutron), which is located within a few femtometers from the nuclear surface of the residual nucleus. Such a configuration is different from the classical description of nuclear reactions at low energies, when all the particles lighter than the residual nucleus in the output channel are assumed to be well separated by the distance from the residual nucleus itself [1]. A possibility for the dineutron to exist as a bound particle was predicted in [2] based on the appearance of the additional bound states of the two neutrons, becoming the dineutron in the external field of the massive nucleus. These bound states correspond to single-particle resonance levels at an additional energy branch which concludes at $\varepsilon_c \sim 0.4$ MeV. Then any single particle states are ranged within [0÷0.4] MeV. The atomic masses of the massive nuclei must be within 100 and 200 µu in order to comprise a system consisting of the heavy nucleus plus the dineutron near the surface of this nucleus [3]. More precise calculations indicate $^{103}$Rh throughout $^{207}$Pb as candidate-nuclei for reaction targets with the dineutron in the output channel. $^{159}$Tb is within this mass region of interest and therefore was selected as a target nuclide in this research. The dineutron was expected to be produced through $(p,^{2}n+n)$ and $(n,^{2}n)$ reactions on $^{159}$Tb over a range of proton and neutron incident energies just below the threshold energy for the corresponding $(p,3n)$ and $(n,2n)$ reactions. The presence of the dineutron in the outgoing channel of these nucleon induced nuclear reactions is greatly facilitated when the residual nucleus is unstable and corresponding irradiations can then be observed. Under such conditions we may consider the new nuclear configuration of the heavy nucleus plus the dineutron, which is trapped at one of the single particle levels. Upon its own decay it may influence some nuclear characteristics of the residual nucleus as the dineutron may not be the only such bound particle in similar nuclear reactions.

*Experimental observations.* - Let's consider two nuclear reactions on $^{159}$Tb in which possible observation of the dineutron could be proved. In [4] the cross-section values are presented for Dy isotopes produced in the proton induced nuclear reactions on $^{159}$Tb from the $(p,n)$ reaction threshold up to 66 MeV. The experimental techniques that were employed for the measurements of the excitation functions are based on the very well-known stacked-foil and sedimented target method. Two stacks consisting of elemental Tb foils and oxide samples, Cu monitor foils, Ti monitor/degrader foils and Al and Cu foils of various thicknesses were used as degraders. The measurement results for $^{159}$Tb $(p,3n)$ $^{157}$Dy excitation function were found to be in "very satisfactory" [4] agreement with ALICE/ASH predictions. One of the very interesting results in the measurement of this reaction cross section is the following: $\sigma_1=1.17\pm0.11$ mb for $E_{p1}=15.47\pm1.78$ MeV while the threshold for this reaction is $E_{th}^{3n} = |-17.14|$ MeV. The upper value of $E_{p1}$ is only 0.11 MeV above the threshold energy and therefore a contribution to this reaction cross section may not only be from the $(p,3n)$ reaction, but also from $(p,^2n+n)$ reaction. Here the energy distribution between the residual nucleus $^{157}$Dy and two more particles (the neutron and the dineutron) can vary considerably. Another measurement of the same reaction cross section is presented in [5]. The stacked-foil technique was employed with application of Tb, Al, La, Al, CeO (sedimented) and Al, which are different stacks of compositions as in [4]. The cross-section measurement results were in a rather good agreement with [4] and "acceptable" [5] with the EMPIRE and ALICE-IPPE predictions. This experimental work provided another extremely interesting result: $\sigma_2$=90±10 µb for $E_{p2}$=14.86±0.85 MeV. In this case the energy of protons is well below the threshold for the $(p,3n)$ reaction. To arrive at a statistical significance of $\sigma_2$ estimate, let's assume that for the "worst case" the total uncertainty in 10 µb includes only a random statistical component due to gamma-rays detection. Then the uncertainty $\Delta S_p$ of a peak area $S_p$ for gamma-ray energy 326.336 keV ($I_\gamma$=93%) is in the relation as follows: $\Delta S_p = S_p/9$. It is well known that $\Delta S_p = \sqrt{S_p + 2 \cdot S_b}$, where $S_b$ is a background area beneath the gamma peak. From this expression we get the following quadratic equation:

$$S_p^2 - 81 \cdot S_p - 162 \cdot S_b = 0.$$

By solving this quadratic equation one can obtain the only positive solution of our interest:

$$S_p = 1/2 \left(81 + \sqrt{6561 + 648 \cdot S_b}\right).$$

Now with application of the following rigid statistical criteria: $S_p = S_b + 5 \cdot \Delta S_p$, the solution can be found: $S_b$=68, $S_p$=153 and $\Delta S_p$=17. Therefore, for any gamma peak area with 153 counts or more the cross-section estimate for $\sigma_2$ is statistically significant. Moreover, if other sources of uncertainty contributed to the total uncertainty of a cross-section value, then this statistical estimate will be even more statistically significant. This means that in the output channel of the $(p,^2n+n)$ reaction the dineutron might be present and located close enough to $^{157}$Dy nuclear surface. Additionally [3] presents results with observation of the $^{158g}$Tb in the outgoing channel of the $^{159}$Tb$(n,^2n)^{158g}$Tb nuclear reaction with an estimated value for this reaction cross section at 75±30 mb for 6.850±0.002 MeV incident neutron energy [6], which is about 1.3 MeV below the corresponding threshold of a $(n,2n)$ nuclear reaction.

*Discussion.* - Based on experimental study of the $^{159}$Tb$(n,^2n)^{158g}$Tb nuclear reaction in [3] the energy range for the binding energy of the dineutron $B_{dn}$ was set as [1.3; 2.8] MeV. However, the proton energy $E_{p2}$ allows for a much better estimate of the lower limit for the binding energy of the dineutron by subtracting 14.86 MeV from the $E_{th}^{3n}$, which gives 2.2 MeV<$B_{dn}$<2.8 MeV. With these new interval limits of the binding energy one can make a new estimate for the half-life of the dineutron ($t_{1/2}$) based on the assumption that the transition $^2n \to d$ occurs via $\beta^-$-decay [7]. Due to low atomic mass of the dineutron ($A_{dn}$=2) this transition was characterized by very low comparative half-life $f_{dn} \cdot t_{1/2}$ value and might occur as superallowed:

$$\lg(f_{dn} \cdot t_{1/2}) = 3 \div 3.5 \text{ with}$$
$$\lg f_{dn} = 4.0 \cdot \lg E_{max-dn} + 0.78 + 0.02 A_d$$
$$- 0.005(A_d - 1) \cdot \lg E_{max-dn}, \quad (1)$$

where $A_d$ - atomic mass of the deuteron, and $E_{max-dn}$ – the limiting kinetic energy of $\beta^-$-decay spectrum in MeV [1]. The new results of dineutron half-life calculations from (1) are presented in Table I resulting in 2.5 s<$t_{1/2}$<20.5 s. By comparing these new half-life limitations to $t_{1/2}$ results as found in [7], one can indicate an acceptable agreement and much less conservative estimates than in [3].

TABLE I. Results of dineutron half-lives calculation.

| $E_{max-dn}$ \ $\lg(f_{dn} \cdot t_{1/2})$ | 2.2 | 2.8 |
|---|---|---|
| 3.0 | 6.5 s | 2.5 s |
| 3.5 | 20.5 s | 7.8 s |

Theoretically if the dineutron is allowed to exist in the configuration of and according to the binding mechanism described above, then other light particles may also behave similarly. As such we cannot exclude and therefore have to reasonably assume that the reaction $^{159}$Tb $(p,3n)$ $^{157}$Dy might not occur only via $(p,3n)$ and $(p,^2n+n)$, but via $(p,^3n)$ channel as well. For the last channel we could estimate the upper limit of the binding energy of the trineutron $^3n$ through again subtracting the energy value 11.16 MeV [5], for which no $^{157}$Dy was detected from $E_{th}^{3n}$, giving an upper limit of 5.88 MeV. From another point of view, should the trineutron exist, it would be susceptible both to neutron radioactivity and $\beta^-$-decay. In the latter case we could assume the following decay channel: $^3n \to {}^3H$ with $\beta^-$ and $\tilde{\nu}_e$ emission. From this decay scheme an upper estimate of the binding energy of the trineutron equals 9.26 MeV. Then the limits for the trineutron binding energy $E_{max-tn}$ would be the following: 2.8 MeV$<B_{tn}<$5.9 MeV. By substituting in (1) $A_d$ for $A_t =3$, $f_{dn}$ for $f_{tn}$, and $t_{1/2}$ for $T_{1/2}$ we get new results of trineutron half-life calculations in Table II.

TABLE II. Results of trineutron half-lives calculation.

| $E_{max-tn}$ \ $\lg(f_{tn} \cdot T_{1/2})$ | 2.8 | 5.9 |
|---|---|---|
| 3.0 | 2.38 s | 0.12 s |
| 3.5 | 7.51 s | 0.38 s |

From Table II we obtain for $T_{1/2}$ [0.12; 7.51] s as an interval estimate of the trineutron half-life. Then experimental observation of the trineutron as a reaction product would be possible for the energies of incident particles $E$ within the following interval: $E_{th}^{3n} - B_{tn} < E < E_{th}^{3n} - B_{dn}$, and, most probably, with application of the particle/neutron activation in-beam technique due to low reaction yield expected with $^3n$ in the output channel. Also, if beta-spectrometry of irradiated Tb sample could be performed, then for the dineutron we may expect detection of an electron spectrum with the limiting energy within [2.2; 2.8] MeV. Similarly, for the trineutron a beta spectrum will be composed of three components: first the limiting energy within [2.2; 2.8] MeV due to the trineutron neutron decay and further dineutron decay into the deuterium; secondly the limiting energy within [2.8; 5.9] MeV due to trineutron $\beta^-$-decay into $^3H$, and finally the limiting energy 18.6 keV due to $^3H$ $\beta^-$-decay into $^3He$. It is fortuitous that all three components are well resolved in time due to significant differences in half lives, where with these above three considerations it would be of some interest to consider the possible influence of dineutron (and, may be, the trineutron) decay products on residual nucleus. Let's take as an example $^{158g}Tb$ in the $(n,2n)$ reaction output channel described above and schematically shown in Fig.1. This nucleus is EC, $\beta^+$- and $\beta^-$-decaying with a 180 year half-life in case the atom of Tb is not affected by any other perturbations. The dineutron is located between $^{158g}Tb$ nucleus and the K-shell of Tb atom and with a half-life of ~ 5-10 s might decay by emitting an electron and $\tilde{\nu}_e$.

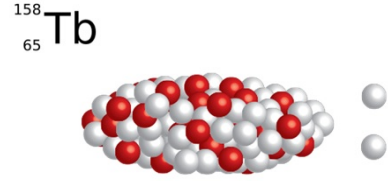

FIG.1. $^{158}Tb$ and the dineutron in the outgoing channel of the $^{159}Tb$ $(n,2n)$ nuclear reaction.

This electron due to the electromagnetic interaction may be attracted and captured by $^{158g}Tb$ with some probability ($P$) followed by the transformation of $^{158g}Tb$ nucleus. Then for the dineutron ($N_{dn}$) and $^{158g}Tb$ ($N_{Tb}$) nuclei the decay rates are expressed with the differential equations system below:

$$\begin{cases} \dfrac{dN_{dn}(t)}{dt} = -\lambda_{dn} \cdot N_{dn}(t) \\ \dfrac{dN_{Tb}(t)}{dt} = -\lambda_{Tb} \cdot N_{Tb}(t) - \lambda_{dn} \cdot P \cdot N_{dn}(t) \end{cases}, \quad (2)$$

where $\lambda_{dn}$ and $\lambda_{Tb}$ are the decay constants of the dineutron and $^{158g}Tb$, correspondingly, and $N_{Tb}(0) = N_{dn}(0) = N_0$. Notice in second equation the first term describes $N_{Tb}$ radioactive decay, while according to the second term $N_{Tb}$ is consumed and transformed into $^{158}Gd$ (stable) because of the electromagnetic interaction between the positively charged nucleus and the electron as a decay product of the dineutron. The solution of the differential equation system (2) is:

$$N_{dn}(t) = N_0 e^{-\lambda_{dn} t},$$
$$N_{Tb}(t) = \frac{P\lambda_{dn}}{\lambda_{dn} - \lambda_{Tb}} N_{dn}(t) + N_0 \frac{\lambda_{dn}(1-P) - \lambda_{Tb}}{\lambda_{dn} - \lambda_{Tb}} e^{-\lambda_{Tb} t}.$$

For the limiting case of $P=1$, $\lambda_{dn} \gg \lambda_{Tb}$ and first 100 s time span since $t=0$ we arrive at the following: $N_{Tb}(t) \approx N_{dn}(t)$. Equivalently the number of nuclei of radioactive $^{158g}Tb$ left in a sample after a certain time interval will be considerably less and determined by dineutron decay constant but not $\lambda_{Tb}$! Luckily, $P \neq 1$ and therefore experiments described above showed some remaining activity for determination of reaction cross sections. Furthermore, it is necessary to keep in mind that under routine practice of cross-section determination, without taking into account a dineutron creation mechanism and a real value of $P$, a cross-section result will be time-dependant, i.e. depending on time period when, for instance, the instrumental gamma spectrum was measured. Dineutrons and trineutrons may have some potential for practical applications where interest could be found in the scenario suggested in [8] for nuclear energy release in metals due to resonant nuclear reactions of low energy dineutrons and trineutrons. Another possible application of the dineutron-nucleus configuration from above would be the treatment of some long-lived radionuclides in radioactive waste management. If the lighter nuclei consisting of two and three neutrons may possibly exist, then why couldn't it be possible to extend this area of research for two and three proton systems, namely the diproton [6] or $^2$He, and the triproton, or $^3$Li (both of them are nuclei, not atoms)? To do so, let's consider the only possible decay channel for a bound diproton: $^2$He $\to$ $^2$H + $\beta^+$ + $\nu_e$, from which follows the upper value of the diproton binding energy: $B_{dp}$ < 0.421 MeV. This energy interval is almost overlapped with an additional energy branch which terminates at $\varepsilon_c \sim 0.4$ MeV with single particle states located within the energy range [0; 0.4] MeV [2]. Considering that for any two nucleons with anti-aligned spins to become bound, at least 0.066 MeV energy [9] must be subtracted, where the limits for diproton binding energy would be the following: 0.066 MeV < $B_{dp}$ < 0.400 MeV. Provided a repulsion energy between the two protons at $r$ (fm) distance is $E_{coul}(r) = 1.44/r$ (MeV), then we can make a radius estimate of the diproton for the binding energy, not exceeding 0.4 MeV, but including 0.066 MeV and a repulsion component. This gives 2.16 fm as a minimal radius of the diproton. Also similar to (1) from [1] for the diproton we have:

$$\lg(f_{dp} \cdot \tau_{1/2}) = 3 \div 3.5 \text{ with}$$

$$\lg f_{dp} = 4.0 \cdot \lg E_{max} + 0.79 - 0.007 A_d$$
$$- 0.009(A_d + 1) \cdot \left[\lg \frac{E_{max}}{3}\right]^3, \quad (3)$$

from where we may get estimates for the diproton half-lives given in Table III.

TABLE III. Results of diproton half-lives calculation.

| $E_{max-dp}$ <br> $\lg(f_{dp} \cdot \tau_{1/2})$ | 0.066 | 0.400 |
|---|---|---|
| 3.0 | 6,650,561 s | 6,286 s |
| 3.5 | 21,030,918 s | 19,846 s |

From Table III we have 6,286 s < $\tau_{1/2}$ < 2.1•10$^7$ s. As such for observation of the diproton it would be worthwhile to irradiate a Tb metal specimen with protons of maximal energy lower then the threshold value of the corresponding (p,2p) reaction. Then the energy interval for incident protons impinging on Tb specimen to observe the diproton should be the following: [5.77; 6.10] MeV. The particle induced gamma-emission in-beam technique might allow detecting gamma transitions between $^{158}$Gd levels and a very intensive annihilation peak due to the diproton decay with half-life within the above estimated time interval. The expected binding energy for the diproton is within [0.37; 0.39] MeV. Likewise, we may assume that the triproton may also exist in the outgoing channel of the nuclear reaction $^{159}$Tb (n,3p). Similarly as the described above, should the triproton exist, it would be susceptible both to proton radioactivity and $\beta^+$–decay. In the latter case the following decay channel looks reasonable: $^3$Li $\to$ $^3$He + $\beta^+$ + $\nu_e$. From this decay scheme an upper estimate of the binding energy of the triproton equals 5.913 MeV and the limits of binding energy $B_{tp}$ for the triproton would be the following: 0.4 MeV < $B_{tp}$ < 5.91 MeV. By substitution in (3) of $A_d$ for $A_{He-3} = 3$, $f_{dp}$ for $f_{tp}$, and $\tau_{1/2}$ for $\check{T}_{1/2}$ we obtain results of triproton half-life calculations, given in Table IV:

TABLE IV. Results of triproton half-lives calculation.

| $E_{max-tp}$ <br> $\lg(f_{tp} \cdot \check{T}_{1/2})$ | 0.4 | 5.91 |
|---|---|---|
| 3.0 | 6,290 s | 0.14 s |
| 3.5 | 19,890 s | 0.44 s |

From Table IV we get 0.14 s < $\check{T}_{1/2}$ < 19,890 s. To observe the triproton it would be necessary to irradiate a Tb metal specimen with neutrons of 14.7 MeV energy and to apply in-beam technique for detecting gammas due to $^{157}$Sm discharge, and a very intensive annihilation peak due to triproton and diproton decay [10]. Additionally, if beta-spectrometry of irradiated Tb sample could be performed, then for the diproton we might expect detection of a positron spectrum with the limiting energy within [0.07; 0.40] MeV. Similarly, for the triproton a $\beta^+$ spectrum will be composed of the two components: the limiting energy within [0.07; 0.4] MeV due to triproton proton decay followed by diproton decay into the deuterium, and secondly the limiting energy within [0.4; 5.91] MeV due to triproton $\beta^+$-decay into $^3$He. In this case the half-lives of both groups overlap to some extent and the annihilation peak will include two additional components. Concluding the binding energy expected for the triproton is within [0.4; 5.91] MeV.

*Summary.* - As it was reasoned in [8] that although there was no direct evidence at that time for the existence of compact, charge-neutral resonant particles like a "virtual" dineutron, "...the indirect evidence for their existence does seem to be increasing". In this letter it was shown that this prediction was rather valid and the dineutron as a bound particle in a close vicinity to the residual nucleus in the outgoing channel of the corresponding nuclear reaction can represent a new nuclear reaction channel. Furthermore, dineutron decay products may cause a significant reduction in the induced activity of the residual nuclei because of interaction of the latter with the electrons due to dineutrons decay. Beside the dineutron, similar reaction and binding mechanisms may lead to appearance of other bound nuclei, composed of the two and even three identical nucleons and known as low nucleon bound systems. Therefore the estimates of dineutron, trineutron, diproton and triproton half-lives and binding energies are given for the first time. While the theoretical description is for now incomplete and based on some experimental facts only, this research may open up a new direction in nuclear physics. More efforts are necessary to shed some light on the process of reliably generating such unique particles as bound dineutrons, trineutrons, diprotons, and triprotons which were historically considered as non-existent. More precise experiments are needed in the future to study these phenomena in details. From what has been stated so far, there may be a chance to fully address the following statement in [2]: "One might think that an analogous mechanism leads to bound states which are more complicated than the dineutron". Finally, one could ask: "Does the dineutron in the outgoing channel considering observation of a bound dineutron represent a new type nuclear reaction on $^{159}$Tb?" This question may be reasonably answered by the following: "Yes, it does. And, probably, not only the dineutron."

*Acknowledgments.* - The author gratefully acknowledges Dr. Nadiia Sakhno for double checking all the results and Dr. Natalia Dzysiuk for pointing out at significant count rates in the annihilation peaks after irradiations of Tb samples with neutrons and protons.